\documentclass[12pt,preprint]{aastex}

\shorttitle{Exoplanet Host Star Companions}
\shortauthors{Baines et al.}

\begin{document}

\title{Ruling Out Possible Secondary Stars to Exoplanet Host Stars Using the CHARA Array}

\author{Ellyn K. Baines$^\dagger$}
\affil{Remote Sensing Division, Naval Research Laboratory, 4555 Overlook Avenue SW, \\ Washington, DC 20375}
\email{ellyn.baines.ctr@nrl.navy.mil}
\altaffiltext{$^\dagger$}{The observations described here were completed while with the Center for High Angular Resolution Astronomy, Georgia State University, P.O. Box 3969, Atlanta, GA 30302-3969.}

\author{Harold A. McAlister, Theo A. ten Brummelaar, Nils H. Turner, Judit Sturmann, Laszlo Sturmann, P. J. Goldfinger, Christopher D. Farrington}
\affil{Center for High Angular Resolution Astronomy, Georgia State University, P.O. Box 3969, Atlanta, GA 30302-3969}
\email{hal@chara.gsu.edu, theo@chara-array.org, nils@chara-array.org, judit@chara-array.org, sturmann@chara-array.org, pj@chara-array.org, farrington@chara-array.org}

\author{Stephen T. Ridgway}
\affil{Kitt Peak National Observatory, National Optical Astronomy Observatory, \\ P.O. Box 26732, Tucson, AZ 85726-6732} 
\email{ridgway@noao.edu}

\altaffiltext{}{For preprints, please email ellyn.baines.ctr@nrl.navy.mil.}

\begin{abstract}
Of the over 450 exoplanets known to date, more than 420 of them have been discovered using radial velocity studies, a method that tells nothing about the inclination of the planet's orbit. Because it is more likely that the companion is a planetary-mass object in a moderate- to high-inclination orbit than a low-mass stellar object in a nearly face-on orbit, the secondary bodies are presumed to be planets. Interferometric observations allow us to inspect the angular diameter fit residuals to calibrated visibilities in order to rule out the possibility of a low-mass stellar companion in a very low-inclination orbit. We used the Center for High Angular Resolution Astronomy (CHARA) Array interferometer to observe 20 exoplanet host stars and considered five potential secondary spectral types: G5 V, K0 V, K5 V, M0 V, and M5 V. If a secondary star is present and is sufficiently bright, the effects of the added light will appear in interferometric observations where the planet will not. All secondary types could be eliminated from consideration for 7 host stars and no secondary stars of any spectral type could be ruled out for 7 more. The remaining 6 host stars showed a range of possible secondary types.
\end{abstract}


\section{Introduction}
Radial velocity observations of exoplanet systems alone are insufficient to distinguish between intermediate- to high-inclination planetary systems and low-inclination binary star systems. \citet{2001A&A...371..250S} estimated probability densities of orbital periods and eccentricities for two samples: exoplanet candidates and spectroscopic binary star systems with solar-type primary stars. They found the distributions of the two populations were statistically indistinguishable in the context of orbital elements.

In an earlier study, \citet{1998A&A...334L..37I} modeled nine exoplanet systems as binary star systems to test if the radial velocity observations could be reproduced by low-mass stellar companions. Although the probability of binary star systems appearing as planetary systems was low -- 0.01 to 4$\%$ -- the model results described the observations satisfactorily and showed it is possible for a binary star system to mimic an exoplanet system.

While it is unlikely that there are unseen stellar companions with very low inclinations masquerading as exoplanets, the only way to exclude the possibility of a low-inclination star observationally is to study the system at angular scales comparable to the calculated star-planet separation. These separations are on the order of 0.5 to 5.0 milliarcseconds, and the only relevant technique applicable is interferometry, which we present here for 20 exoplanet host stars.

The probability of a system's inclination being in the range $i$ to $i + \Delta i$ is proportional to the ratio of the surface element of a hemisphere defined by that range and integrated over the azimuth angle (\emph{$\Phi$}) to the surface area of the entire hemisphere. The area element for a given range of \emph{i} is: 
\begin{equation} 
\mathrm{d} A = \mathrm{d} i \times \mathrm{sin} i \; \mathrm{d} \Phi,
\end{equation}
and the probability of a system having a specific range of \emph{i} is:
\begin{equation}
P_{i,i+ \Delta i} = \frac{\int^{2 \pi}_0 \int^{i+ \Delta i}_i \sin i \; \mathrm{d}i \; \mathrm{d} \Phi}
{\int^{2 \pi}_0 \int^{\pi / 2}_0 \sin i \; \mathrm{d} i \; \mathrm{d} \Phi} 
= \frac{-2 \pi \; \cos i \mid^{i+ \Delta i}_i}{-2 \pi \; \cos i \mid^{\pi / 2}_0} 
= \cos i - \cos (i+ \Delta i).
\end{equation}
Therefore the probability of an orbit with an inclination below 45$^\circ$ is $\sim$30$\%$ while the probability of the orbit having an inclination higher than 45$^\circ$ is $\sim$70$\%$. The inclinations necessary for the companion to be stellar in nature are very low, on the order of less than one degree. This means the probability of any given system being a binary star is correspondingly low ($\sim 10^{-4} \%$) and it would take much larger sample size than is available to have a good chance of finding a stellar companion masquerading as a planet. When taken as a whole, the chances of finding a stellar companion in this sample are only on the order of $10^{-5}$.

We are not trying to prove that the apparent companion for any given system is a low-mass star instead of a planet, but instead are using our observations to rule out certain types of stellar companions for each star observed. Radial velocity studies alone cannot perform this task, and interferometry is well suited to further our knowledge of exoplanet host stars.  When we are able to be more confident that they are indeed planetary systems and not face-on binary stars, we further characterize the host stars themselves as well as contribute to the statistics that tell us what percentage of stars host planets versus how many have stellar companions.

An important criteria for the detection or non-detection of otherwise unseen secondary stars is the magnitude difference between the known primary and the putative secondary stars. Simulations using a program written by Theo ten Brummelaar that realistically models instrumental and atmospheric noises, as well as observations of pairs of known brightness contrasts, indicate that the Array is sensitive to a magnitude difference in the $K$-band ($\Delta K$) of 3.0. Therefore, if a second star is present and is not more than $\sim$3.0 magnitudes fainter than the host star, the effects of the second star will be seen in the interferometric data. It should be noted that a limiting magnitude difference in the $K$-band ($\Delta K$) of 3.0 is a lower limit, as the true $\Delta K$ also depends on the absolute brightness of the two stars and could be slightly higher for some systems.

This technique of using interferometric observations to eliminate the possibility of certain types of secondary stars was employed to examine the exoplanet host star 51 Peg (HD~217014) by \citet{1998ApJ...504L..39B}, whose analysis of Palomar Testbed Interferometer data supported a single-star model for that star. They fit single-star and binary-star models to the data and found that any possible unseen stellar companion would have to have a $K$ magnitude fainter than 7.30 and a mass of less than 0.22$M_\odot$.

Here we describe our interferometric observations, our method for choosing calibrator stars, and define the role interferometric resolution plays in Section 2. In Section 3, we discuss how the angular diameter fit residuals to calibrated visibilities can help us eliminate certain types of secondary stars, and Section 4 explores the implications of the observations. This paper is follow-on work to an earlier study \citep{2008ApJ...682..577B}.


\section{Interferometric Observations}
All observations were obtained using the CHARA Array, a six-element optical/infrared interferometric array located on Mount Wilson, California \citep{2005ApJ...628..453T}. We used the pupil-plane ``CHARA Classic'' beam combiner in the $K'$-band (2.133 $\mu$m center with a 0.349 $\mu$m width) while visible wavelengths (470-800 nm) were used for tracking and tip/tilt corrections. The observing procedure and data reduction process employed here are described in \citet{2005ApJ...628..439M}. The observable quantity from an interferometer is the fringe contrast or ``visibility'' of the observed target, and each dataset consists of approximately 200 scans across the fringe.

Our target list was selected from the complete exoplanet list by using declination limits and magnitude constraints: north of -10$^\circ$ declination, brighter than $V=+10$ in order for the tip/tilt system to lock onto the star, and brighter than $K=+6.5$ for reliable fringe detection with a sufficiently high signal-to-noise ratio. We obtained data on the 20 exoplanet host stars between October 2005 and September 2008. The observations were taken using mostly the longest baseline available on the CHARA Array (331 m), though 156-m and 249-m baselines were also used.

Reliable calibrators stars are critical in interferometric observations, acting as the standard against which the science target is measured, and the ideal calibrator is a single, spherical, non-variable star. Our observing pattern was calibrator-target-calibrator so that every target was bracketed by calibrator observations made as close in time as possible; therefore ``5 bracketed observations'' denotes 5 target and 6 calibrator data sets. The target-calibrator (T-C) distances ranged from 1 to 9$^\circ$ and 13 calibrators were within 4$^\circ$ of their target stars. This allowed us to observe the stars as close together in time as possible, usually on the order of 3 to 5 minutes between the two, therefore reducing the effects of changing seeing conditions as much as possible. Table \ref{observations} lists the exoplanet host stars observed, their calibrators, the dates of the observations, the baseline used, the number of observations obtained, and the T-C distance.

In order to check for excess emission that could indicate a low-mass stellar companion or circumstellar disk, we fitted spectral energy distributions (SEDs) based on published $UBVRIJHK$ photometric values for each calibrator star. Limb-darkened diameters were calculated using Kurucz model atmospheres\footnote{See http://kurucz.cfa.harvard.edu.} based on effective temperature and gravity values obtained from the literature. The models were then fit to observed photometric values also from the literature after converting magnitudes to fluxes using \citet{1996AJ....112..307C} for $UBVRI$ values and \citet{2003AJ....126.1090C} for $JHK$ values. 

Many of the calibrator stars chosen here had been used as comparison or calibrator stars in other studies, or speckle studies did not find companions (see Table \ref{calib_info}). For those calibrator stars that had not been previously observed, their SED fits showed no excess flux that could indicate a stellar companion that would then contaminate our interferometric observations.

\begin{figure}[!t]
  \centering \includegraphics[angle=90,width=1.0\textwidth]
  {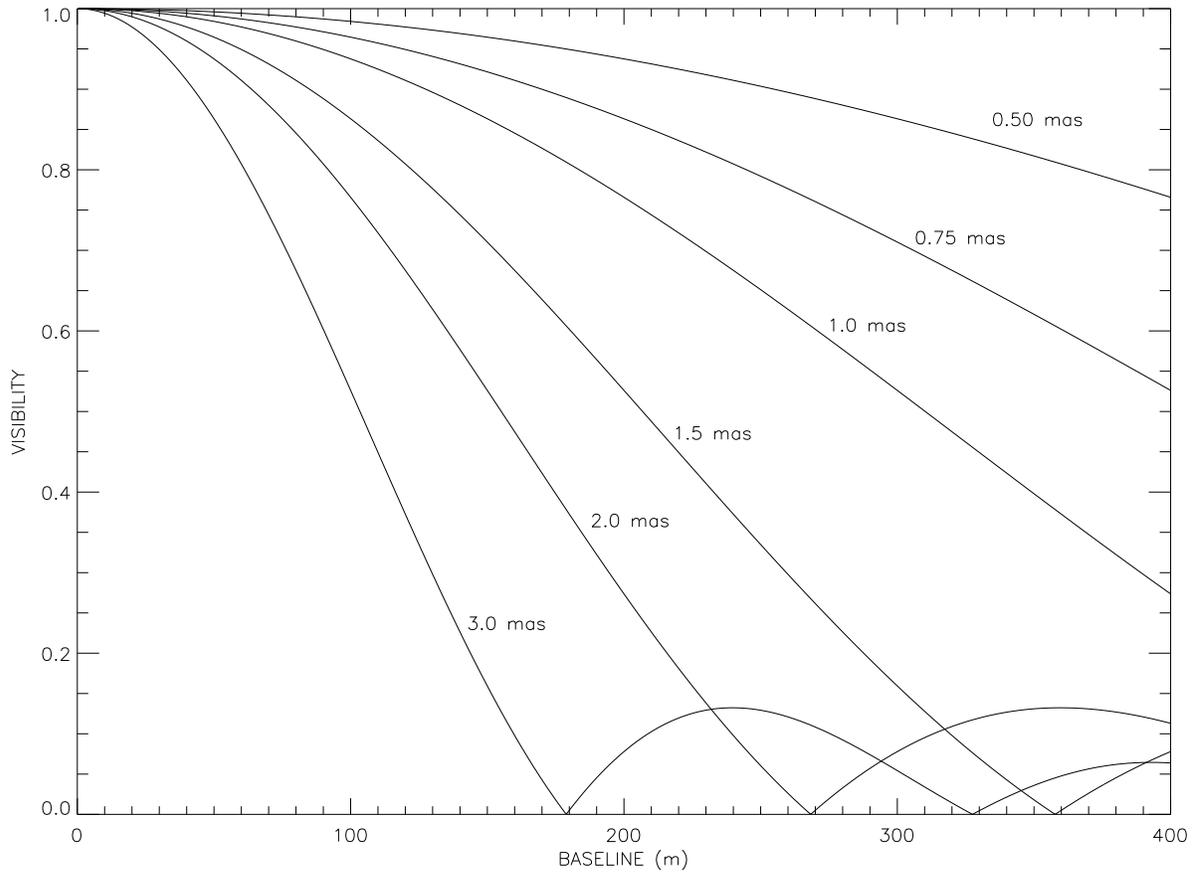}\\
 \caption{The effect of various stellar angular diameters on the visibility curve. The smaller a star's angular diameter, the less change is seen in the visibility curve as a function of baseline.}
  \label{angdiam}
\end{figure}

Our ability to detect stellar companions depends on two main factors. The first is the precision of our visibility measurements. The higher the precision, the higher our sensitivity to finding a secondary companion. The second factor is whether the measured angular diameters or potential primary-secondary separation would be resolved in our data. The resolution of an interferometer depends on the wavelength used and the distance between the telescopes, otherwise known as the baseline. A star is considered unresolved if its visibility is $\cong$1 and is completely resolved when its visibilities drop to zero. Different sized stars will be resolved at different baselines (see Figure \ref{angdiam}). 

Another effect to account for is bandwidth smearing, which occurs when the physical width of the filter's bandpass affects the measurements as the resolution varies across the band. Bandwidth smearing is only significant when a star's angular diameter exceeds the coherent field of view of the interferometer, which is calculated to be
\begin{equation}
\mathrm{FOV} = \frac{\theta_{\rm min}}{\pi} \left( \frac{\Delta \lambda}{\lambda_0} \right)^{-1},
\end{equation}
where $\theta_{\rm min} = \lambda_0 / B$ and $B$ is the baseline, $\lambda_0$ is the central wavelength of the filter, and $\Delta \lambda$ is the width of the filter \citep{2002MNRAS.333..642T}. Because our coherent FOV is larger than the measured angular diameters in all cases, we do not need to correct for this effect when measuring our primary exoplanet host stars. On the other hand, when we determine the visibilities for binary systems that have calculated separations larger than the FOV, we need to account for bandwidth smearing by using a modified version of the visibility equation for binary stars.


\section{Characterizing Angular Diameter Fit Residuals to Calibrated Visibilities}
\begin{figure}[!t]
  \centering \includegraphics[angle=90,width=1.0\textwidth]
  {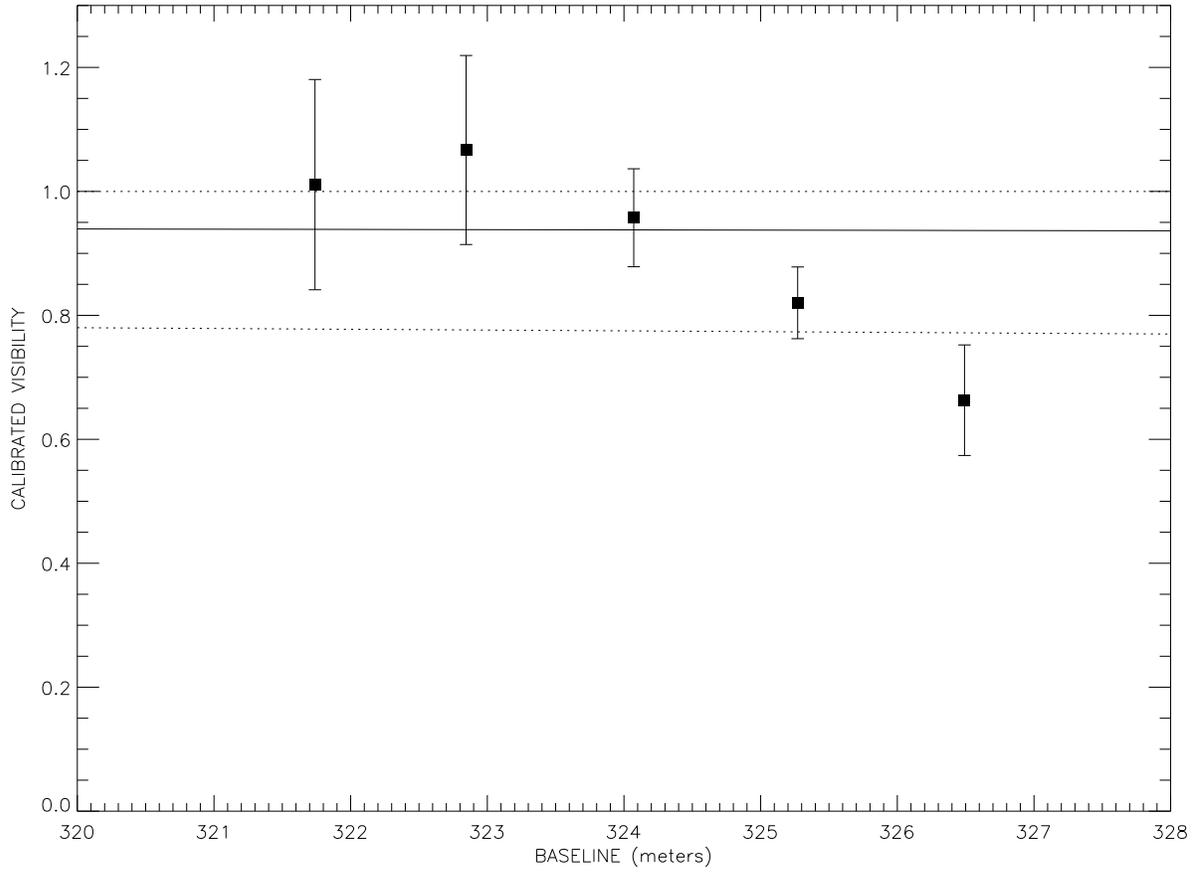}\\
 \caption{HD 164922 LD disk diameter fit. The solid line represents the theoretical visibility curve for the star with the best fit $\theta_{\rm LD}$, the dashed lines are the 1$\sigma$ error limits of the diameter fit, the squares are the calibrated visibilities, and the vertical lines are the measured errors.}
  \label{HD164922_lddiam}
\end{figure}

To determine stellar angular diameters, measured visibilities ($V$) are fit to a model of a uniformly-illuminated disk (UD). Single-star diameter fits to $V$ were based upon the UD approximation given by $V = [2 J_1(x)] / x$, where $J_1$ is the first-order Bessel function and $x = \pi B \theta_{\rm UD} \lambda^{-1}$, where $B$ is the projected baseline at the star's position, $\theta_{\rm UD}$ is the apparent UD angular diameter of the star, and $\lambda$ is the effective wavelength of the observation \citep{shao92}. A more realistic model of a star's disk involves limb-darkening (LD), and the relationship incorporating the linear limb darkening coefficient $\mu_{\lambda}$ \citep{han74} is:
\begin{equation}
V = \left( {1-\mu_\lambda \over 2} + {\mu_\lambda \over 3} \right)^{-1}
\times
\left[(1-\mu_\lambda) {J_1(\rm x) \over \rm x} + \mu_\lambda {\left( \frac{\pi}{2} \right)^{1/2} \frac{J_{3/2}(\rm x)}{\rm x^{3/2}}} \right] .
\end{equation}
Table \ref{calib_visy} lists the Modified Julian Date (MJD), baseline $B$, projected baseline position angle ($\Theta$), calibrated visibility ($V_c$), and error in $V_c$ ($\sigma V_{c}$) for each star observed, and the resulting angular diameters are presented in Table \ref{host_obs_props}. Figure \ref{HD164922_lddiam} shows the LD diameter fit for HD 164922 as this star's diameter is presented for the first time here. Similar plots for the remaining stars can be found in the references listed in Table \ref{host_obs_props}. 

The systematics in the residuals of the angular diameter fit to measured visibilities can help us eliminate certain types of potential secondary stars. The smaller the residuals, the lower the chance of an unseen stellar companion. For each exoplanet host star observed, a variety of secondary stars were considered: G5~V, K0~V, K5~V, M0~V, and M5~V. The magnitude difference ($\Delta$M$_K$, listed as $\Delta K$ in the tables) and angular separation ($\alpha$) of a face-on orbit between the host star and companion were calculated for each possible pairing:
\begin{equation}
\Delta \mathrm{M}_K = \mathrm{M}_{\rm s} + (\mathrm{m}_{\rm h} - \mathrm{M}_{\rm h}) - \mathrm{m}_{\rm h},
\end{equation}
where M$_{\rm h,s}$ are the absolute magnitudes of the host star and potential secondary, respectively, m$_{\rm h}$ is the apparent magnitude of the host star, and
\begin{equation}
\rm (m_{h} - M_{h}) = 5 \; log \left( \frac{100}{\pi} \right),
\end{equation}
where $\pi$ is the host star's parallax in milliarcseconds (mas). An estimate of the angular separation $\alpha$ in mas was calculated from Kepler's Third Law:
\begin{equation}
\alpha = \left[ \left(M_{\rm h}+M_{\rm s} \right) \times P^2 \right]^{\frac{1}{3}} \times \pi ,
\end{equation}
where $M_{\rm h,s}$ are the masses in $M_\odot$ of the exoplanet's host star and potential secondary star, respectively, and $P$ is the companion's orbital period in years.

The angular diameter ($\theta$) for each possible secondary star was estimated using the calibration of radius as a function of spectral type from \citet{2000asqu.book.....C} and the parallax of the host star. The masses and radii in Cox are based on values derived from \citet{1981A&AS...46..193H}, who observed binary stars in order to create empirical relationships between various stellar parameters as a function of luminosity class. Tables \ref{host_obs_props} and \ref{poss_sec_params} present the results of these calculations. 

The resulting values for $\theta$, $\Delta K$, and $\alpha$ were then used to determine the visibility curve for a single star with the host star's measured angular diameter as well as for a binary system with the parameters listed above. The equation used to calculate the visibility curve for a binary system when observed using a narrow bandpass is 
\begin{equation}
V = (1 + \beta)^{-1} [V_1^2 + \beta^2 V_2^2 + 2 \beta V_1 V_2 \cos \{2 \pi B \lambda^{-1} \alpha \cos \phi\} ]^{\frac{1}{2}},
\end{equation}
where $V_{1,2}$ are the visibilities for the primary and secondary star, respectively,  $\beta = 100^{0.2 \Delta \rm m}$ where $\Delta \rm m$ is the magnitude difference between the two stars, and $\phi$ is the difference of the position angles between the binary and baseline \citep{1970MNRAS.148..103H}. Because the observations described here were taken using a filter with a bandpass of $\sim 16 \%$, the equation needs to be modified to include the effects of wide bandwidth and bandwidth smearing \citep{2007MNRAS.377..415N}: 
\begin{equation}
V = (1 + \beta)^{-1} [\beta^2 V_1^2 + V_2^2 + 2 \beta r(\psi) V_1 V_2 \cos \psi ]^{\frac{1}{2}},
\end{equation}
where $\psi = 2 \pi B \lambda^{-1} \alpha \cos \phi$ and
\begin{equation}
r(\psi) = \exp \left( \frac{- \Delta \lambda^2}{\lambda^2_0} \frac{\psi^2}{32 \ln 2} \right).
\end{equation}
To explore the effects of the projected position angle of a binary star vector separation, we calculated the residuals using position angles of 0$^\circ$, 30$^\circ$, and 60$^\circ$ (see Table \ref{calc_resids}). 

\begin{figure}[!t]
  \centering \includegraphics[angle=90,width=1.0\textwidth]
  {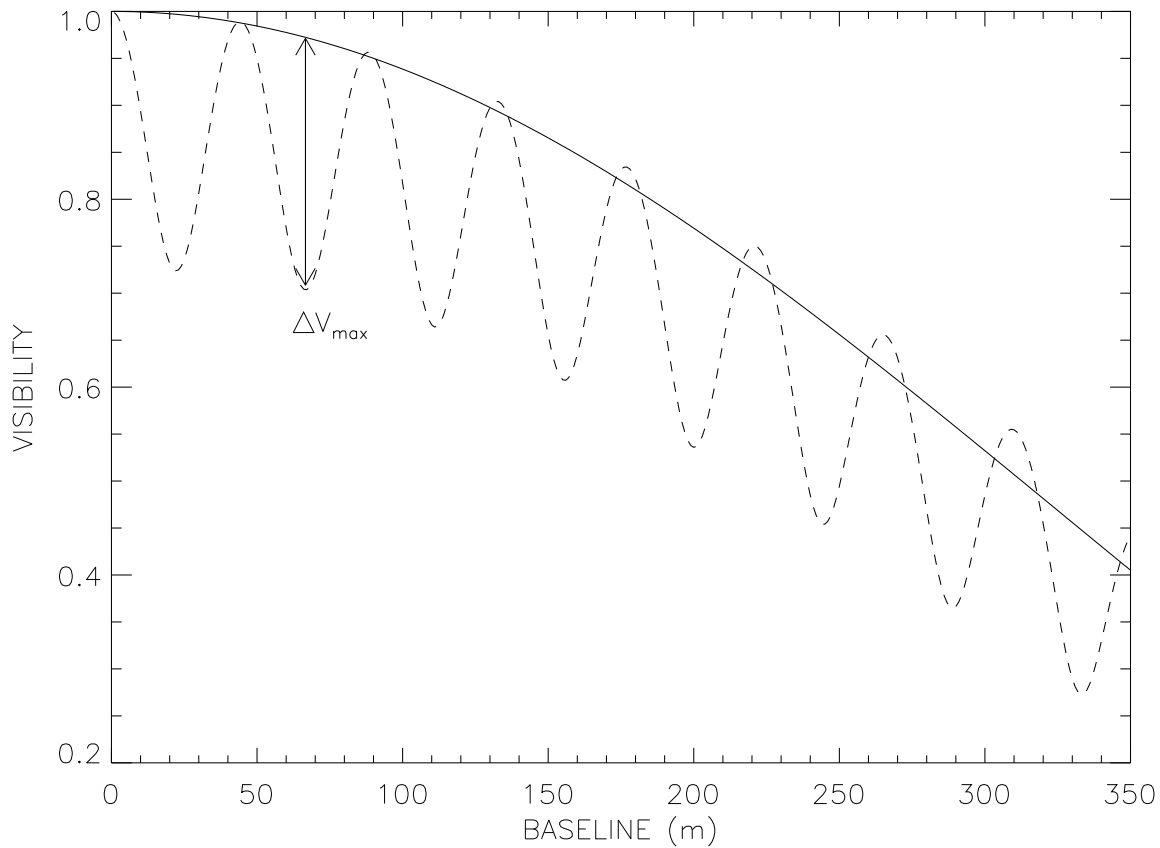}\\
 \caption{Example of the difference between the visibility curves for a single star and a binary system. The solid line indicates the curve for a single star with $\theta$=1.0~mas, while the dashed line represents the curve for a binary system with the following parameters: $\theta_{\rm primary}$=1.0~mas, $\theta_{\rm secondary}$=0.5~mas, $\alpha$=10~mas, and $\Delta K$=2.0. $\Delta V_{\rm max}$ is the maximum deviation between the two curves.}
  \label{singlevsbinary_viscurve}
\end{figure}

To estimate the detection sensitivity, the largest difference between the visibility curves for a single star and for a binary system with the parameters listed in Table \ref{poss_sec_params} was calculated. This quantity, $\Delta V_{\rm max}$, then represented the maximum deviation of the binary visibility curve from the single-star curve. Figure~\ref{singlevsbinary_viscurve} shows an example of this.

Due to uncertainties in such input parameters as the host star's mass, parallax, and the planet's orbital period used in Equations 3 and 4, we did not believe a 1$\sigma$ threshold would be a reliable diagnostic. Therefore, in order to rule out putative stellar companions, we selected a lower limit of 2$\sigma_{\rm res}$, where $\sigma_{\rm res}$ is the standard deviation of the residuals to the diameter fit; i.e., if $\Delta V_{\rm max} \geq 2 \sigma_{\rm res}$ for a given secondary component, that particular spectral type can be eliminated as a possible stellar companion. If $\Delta V_{\rm max} < 2 \sigma_{\rm res}$, the effects of the companion would not be clearly seen in the visibility curve, and that spectral type cannot be ruled out. For each exoplanet host star, Table \ref{calc_resids} lists the observed $\sigma_{\rm res}$ and the predicted $\Delta V_{\rm max}$ for each secondary type considered.


\section{Results and Discussion}
Though the errors for the individual calibrated visibilities listed in Table \ref{calib_visy} are on the order of 2-20$\%$, the errors for the angular diameter fits to these visibilities are on the order of 1-6$\%$ for most stars. This is because the calibrated visibility errors are overestimated. This can be best illustrated by performing the diameter fit to the data. For each $\theta_{\rm LD}$ fit, the errors were derived via the reduced $\chi^2$ minimization method \citep{2003psa..book.....W,1992nrca.book.....P}: the diameter fit with the lowest $\chi^2$ was found and the corresponding diameter was the final $\theta_{\rm LD}$ for the star. The errors were calculated by finding the diameter at $\chi^2 + 1$ on either side of the minimum $\chi^2$ and determining the difference between the $\chi^2$ diameter and $\chi^2 +1$ diameter. In calculating the diameter errors in Table~\ref{host_obs_props}, we adjusted the estimated visibility errors to force the reduced $\chi^2$ to unity because when this is omitted, the reduced $\chi^2$ is well under 1.0, indicating we are overestimating the errors in our calibrated visibilities.

\begin{figure}[!h]
  \centering \includegraphics[width=0.5\textwidth]
  {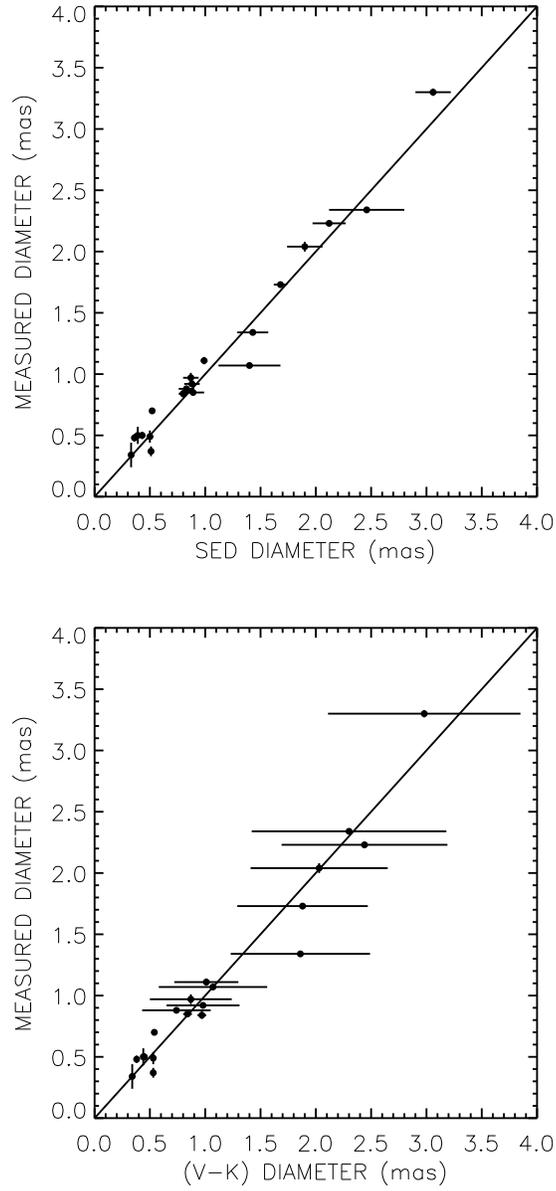}\\
  \caption{A comparison of estimated and interferometrically-measured angular diameters. The upper and lower panels compare diameters derived using SED fits and ($V-K$) colors, respectively, versus diameters measured using the CHARA Array. Note the larger error bars associated with the SED and ($V-K$) diameters for stars larger than $\sim$0.7 mas.}
  \label{diam_compare}
\end{figure}

In addition to measuring the angular diameters of these stars interferometrically, we also estimated their diameters using two methods to check for discrepancies. We performed SED fits using the method described in Section 2 as well as using the relationship described in \citet{2004AandA...426..297K} between the ($V-K$) color and log $\theta_{\rm LD}$. Table \ref{ld_sed_vk} lists the results of these calculations and Figure \ref{diam_compare} plots $\theta_{\rm SED}$ and $\theta_{(V-K)}$ versus $\theta_{\rm measured}$. For stars larger than $\sim$0.7 mas, the errors in the estimated diameters are larger than those for the measured diameters. 

In order to characterize the scatter in the diameters of the entire sample, the standard deviation $\sigma$ of the quantity $| \theta_{\rm LD} - \theta_{\rm SED} |$ was determined to be $8\%$, which indicates a fairly good correspondence between the estimated and measured diameters. For comparison purposes, the standard deviation of $| \theta_{(V-K)} - \theta_{\rm SED} |$ was $12\%$. Four of the 20 stars in the sample have measured diameters that are not within 1-sigma of the diameters estimated using either SED fits or ($V-K$) color. Three of the four (HD~145675, HD~154345, and HD~185269) are among the smallest stars measured and have some of the highest diameter errors in the sample, ranging from 6$\%$ to 11$\%$. 

The largest outlier is HD~217107, which we measured at 0.70$\pm$0.01 mas while the diameters from SED fits and ($V-K$) color were 0.52$\pm$0.02 mas and 0.54$\pm$0.02 mas, respectively. There are no signs of variability indicated in the literature for the star that would impact the diameters estimated using photometry. While the calibrator was small (0.31$\pm$0.01 mas), showed no signs of having a stellar companion using speckle \citep{1987AJ.....93..183M} or in the SED fit, and was used as a photometric comparison star for HD~217107 \citep{2005ApJ...632..638V}, it could be the cause of the discrepancy in the angular diameter estimates and interferometric measurements. Future observations of HD~217107 using the CHARA Array and different calibrators should help clarify the situation.

The star that showed the most potential for being a binary system instead of a planetary system was the newly-presented HD~164922. The visibility points show a slight sinusoidal pattern, though there are not enough observations to reliably fit the data to a binary star model. Future planned observations using the CHARA Array over a longer time should provide more details on the star.

No secondary spectral types could be eliminated from consideration for 7 exoplanet hosts, while all spectral types could be discounted for 7 host stars. The remaining 6 host stars had some but not all of the various secondary types ruled out. Because of the small sample size, we did not expect to find any stellar companions masquerading as planets, as the probability of a moderate- to high-inclination planet mimicking a face-on stellar companion is very low. Our contribution was eliminate the possibility of certain secondary spectral types for the host stars. 

\acknowledgments

The CHARA Array is funded by the National Science Foundation through NSF grant AST-0908253 and by Georgia State University through the College of Arts and Sciences, and STR acknowledges partial support by NASA grant NNH09AK731. This research has made use of the SIMBAD literature database, operated at CDS, Strasbourg, France, and of NASA's Astrophysics Data System. This publication makes use of data products from the Two Micron All Sky Survey, which is a joint project of the University of Massachusetts and the Infrared Processing and Analysis Center/California Institute of Technology, funded by the National Aeronautics and Space Administration and the National Science Foundation.

\clearpage

\begin{deluxetable}{ccclcc}
\tablewidth{0pc}
\tabletypesize{\scriptsize}
\tablecaption{Observing Log.\label{observations}}

\tablehead{
 \colhead{Target} & \colhead{Calibrator} & \colhead{Baseline}      & \colhead{Date} & \colhead{\#}  & \colhead{T-C Sep} \\
 \colhead{HD}     & \colhead{HD}         & \colhead{(max. length)} & \colhead{(UT)} & \colhead{Obs} & \colhead{(deg)} \\ }
\startdata
10697  & 10477  & S1-E1 (331 m) & 2005 Oct 23 & 4  & 4 \\
       &        &               & 2007 Sep 14 & 4  &   \\
13189  & 11007  & S1-E1 (331 m) & 2005 Dec 12 & 4  & 4 \\
       &        &               & 2006 Aug 14 & 4  &   \\
32518  & 31675  & S1-E1 (331 m) & 2007 Nov 14 & 9  & 3 \\
45410  & 46590  & S1-E1 (331 m) & 2008 Sep 11 & 5  & 2 \\
50554  & 49736  & S1-E1 (331 m) & 2005 Dec 12 & 5  & 2 \\
73108  & 69548  & E2-W2 (156 m) & 2008 May 9  & 5  & 7 \\
136726 & 145454 & E2-W2 (156 m) & 2008 May 9  & 6  & 6 \\
139357 & 132254 & S1-E1 (331 m) & 2007 Sep 14 & 4  & 7 \\
145675 & 151044 & S1-E1 (331 m) & 2006 Aug 12 & 6  & 8 \\
154345 & 151044 & S1-E1 (331 m) & 2008 Sep 10 & 7  & 4 \\
164922 & 159139 & S1-E1 (331 m) & 2008 Aug 11 & 5  & 7 \\
167042 & 161693 & S1-E1 (331 m) & 2007 Sep 15 & 8  & 4 \\
170693 & 172569 & W1-S2 (249 m) & 2007 Sep 3  & 4  & 1 \\
185269 & 184381 & S1-E1 (331 m) & 2008 Jul 18 & 15 & 3 \\
       &        &               & 2008 Jul 20 & 5  &   \\
188310 & 182101 & S1-E1 (331 m) & 2008 Sep 8  & 8  & 8 \\
199665 & 194012 & S1-E1 (331 m) & 2008 Sep 8  & 10 & 9 \\
210702 & 210074 & S1-E1 (331 m) & 2008 Sep 8  & 4  & 4 \\
217107 & 217131 & S1-E1 (331 m) & 2008 Sep 8  & 5  & 1 \\
221345 & 222451 & S1-E1 (331 m) & 2008 Sep 11 & 5  & 3 \\
222404 & 219485 & S1-E1 (331 m) & 2008 Sep 11 & 7  & 4 \\
\enddata
\tablecomments{The three arms of the Array are denoted by their cardinal directions: ``S'' is south, ``E'' is east, and ``W'' is west. Each arm bears two telescopes, numbered ``1'' for the telescope farthest from the beam combining laboratory and ``2'' for the telescope closer to the lab.
}
\end{deluxetable}

\clearpage


\begin{deluxetable}{cl}
\tablewidth{0pc}
\tablecaption{Notes on Calibrator Quality and Previous Uses.\label{calib_info}}
\tablehead{\colhead{HD} & \colhead{ } }
\startdata
10477  & No sign of duplicity in literature or SED fit \\
\hline
11007  & Listed as a ``suitable'' calibrator in \citet{2008ApJS..176..276V}; \\
       & used as calibrator in \citet{2004ApJ...610..443K} \\
\hline
31675  & No companion found using speckle in \citet{1989AJ.....97..510M} \\
\hline
46590  & No sign of duplicity in literature or SED fit \\
\hline
49736  & No sign of duplicity in literature or SED fit \\
\hline
69548  & No sign of duplicity in literature or SED fit \\
\hline
132254 & Listed as a ``probably suitable'' calibrator in \citet{2008ApJS..176..276V}; \\
       & no companion found using speckle in \citet{1989AJ.....97..510M} \\
\hline
145454 & No sign of duplicity in literature or SED fit \\
\hline
151044 & Listed as a ``probably suitable'' calibrator in \citet{2008ApJS..176..276V} \\
\hline
159139 & No companion found using speckle in \citet{1987AJ.....93..183M} \\
\hline
161693 & No companion found using speckle in \citet{1987AJ.....93..183M} \\
\hline
172569 & Listed in ``HIP Visual Binaries Kinematics'' table by \citet{2001BaltA..10..481B}; \\
       & but no other information given in paper or general literature \\
\hline
182101 & Used as calibrator in \citet{2006ApJ...644..475B} \\
\hline
184381 & Used as comparison star in \citet{2006ApJ...652.1724J} \\
\hline
194012 & Listed as a ``suitable'' calibrator in \citet{2008ApJS..176..276V}; \\
       & no companion found using speckle in \citet{1987AJ.....93..183M} \\
\hline
210074 & Listed as a ``suitable'' calibrator in \citet{2008ApJS..176..276V}; \\
       & used as comparison star in \citet{2005ApJ...632.1157W} \\
\hline
217131 & No companion found using speckle in \citet{1987AJ.....93..183M} \\
\hline
219485 & No sign of duplicity in literature or SED fit \\
\hline
222451 & No sign of duplicity in literature or SED fit \\
\enddata
\end{deluxetable}

\clearpage


\begin{deluxetable}{ccccccc}
\tablewidth{0pc}
\tabletypesize{\scriptsize}
\tablecaption{Exoplanet Host Stars' Calibrated Visibilities.\label{calib_visy}}
\tablehead{
 \colhead{Target} & \colhead{ }   & \colhead{$B$} & \colhead{$\Theta$}    & \colhead{ } & \colhead{ } & \colhead{$\%$} \\
 \colhead{Name}     & \colhead{MJD} & \colhead{(m)}      & \colhead{(deg)} & \colhead{$V_c$} & \colhead{$\sigma V_c$} & \colhead{err} \\ }
\startdata
10697 & 53666.427 & 309.50 & 175.2 & 1.043 & 0.090 & 9 \\
      & 53666.443 & 310.79 & 171.2 & 0.928 & 0.068 & 7 \\
      & 53666.456 & 312.37 & 167.9 & 1.004 & 0.104 & 10 \\
      & 53666.470 & 314.46 & 164.6 & 0.832 & 0.075 & 9 \\
      & 54356.765 & 317.17 & 226.9 & 0.883 & 0.103 & 12 \\
      & 54356.775 & 320.54 & 227.4 & 1.029 & 0.125 & 12 \\
      & 54357.783 & 323.64 & 228.1 & 0.851 & 0.077 & 9 \\
      & 54357.792 & 325.84 & 228.8 & 0.723 & 0.068 & 9 \\
      & 54357.802 & 327.87 & 229.7 & 0.816 & 0.089 & 11 \\
      & 54357.810 & 328.95 & 230.4 & 0.791 & 0.061 & 8 \\
13189 & 53716.270 & 327.09 & 184.4 & 0.607 & 0.056 & 9 \\
      & 53716.285 & 326.91 & 180.9 & 0.531 & 0.081 & 15 \\
      & 53716.298 & 326.96 & 177.7 & 0.589 & 0.095 & 16 \\
      & 53716.312 & 327.21 & 174.3 & 0.575 & 0.130 & 23 \\
      & 53961.441 & 326.60 & 216.4 & 0.622 & 0.051 & 8 \\
      & 53961.454 & 328.41 & 214.4 & 0.648 & 0.062 & 10 \\
      & 53961.467 & 329.65 & 212.2 & 0.643 & 0.073 & 11 \\
      & 53961.481 & 330.38 & 209.8 & 0.607 & 0.040 & 7 \\
32518 & 54418.238 & 230.84 & 200.1 & 0.755 & 0.067 & 9 \\
      & 54418.244 & 233.56 & 201.8 & 0.794 & 0.071 & 9 \\
      & 54418.250 & 236.48 & 203.6 & 0.834 & 0.070 & 8 \\
      & 54418.256 & 239.18 & 205.3 & 0.843 & 0.074 & 9 \\
      & 54418.261 & 241.66 & 206.9 & 0.751 & 0.061 & 8 \\
      & 54418.267 & 244.20 & 208.6 & 0.743 & 0.053 & 7 \\
      & 54418.274 & 246.86 & 210.3 & 0.776 & 0.059 & 8 \\
      & 54418.280 & 249.36 & 212.0 & 0.741 & 0.065 & 9 \\
      & 54418.286 & 251.81 & 213.8 & 0.732 & 0.053 & 7 \\
45410 & 54720.481 & 258.11 & 212.9 & 0.696 & 0.078 & 11 \\
      & 54720.490 & 263.24 & 215.0 & 0.651 & 0.053 & 8 \\
      & 54720.496 & 266.69 & 216.4 & 0.587 & 0.073 & 12 \\
      & 54720.502 & 269.68 & 217.7 & 0.665 & 0.106 & 16 \\
      & 54720.509 & 272.90 & 219.2 & 0.716 & 0.097 & 14 \\
50554 & 53711.523 & 317.33 & 174.2 & 0.874 & 0.127 & 15 \\
      & 53711.537 & 318.32 & 170.7 & 0.783 & 0.090 & 11 \\
      & 53716.422 & 321.04 & 195.4 & 1.006 & 0.138 & 14 \\
      & 53716.435 & 319.48 & 192.2 & 0.905 & 0.091 & 10 \\
      & 53716.449 & 318.20 & 189.0 & 0.984 & 0.083 & 8 \\
      & 53716.463 & 317.30 & 185.7 & 1.027 & 0.096 & 9 \\
      & 53716.479 & 316.73 & 181.7 & 0.956 & 0.150 & 16 \\
73108 & 54595.216 & 155.95 & 254.7 & 0.411 & 0.051 & 12 \\
      & 54595.226 & 155.88 & 258.0 & 0.446 & 0.034 & 8 \\
      & 54595.235 & 155.83 & 261.1 & 0.436 & 0.043 & 10 \\
      & 54595.244 & 155.80 & 264.1 & 0.460 & 0.057 & 12 \\
      & 54595.257 & 155.77 & 268.4 & 0.430 & 0.092 & 21 \\
136726 & 54595.294 & 147.57 & 189.4 & 0.442 & 0.055 & 12 \\
       & 54595.307 & 148.79 & 193.7 & 0.425 & 0.045 & 11 \\
       & 54595.315 & 149.53 & 196.5 & 0.468 & 0.054 & 12 \\
       & 54595.325 & 150.30 & 199.6 & 0.421 & 0.056 & 13 \\
       & 54595.336 & 151.17 & 203.4 & 0.436 & 0.062 & 14 \\
       & 54595.346 & 151.80 & 206.5 & 0.409 & 0.053 & 13 \\
139357 & 54357.149 & 320.57 & 102.8 & 0.450 & 0.070 & 16 \\
       & 54357.155 & 320.14 & 104.2 & 0.460 & 0.045 & 10 \\
       & 54357.161 & 319.66 & 105.6 & 0.487 & 0.063 & 13 \\
       & 54357.167 & 319.12 & 107.1 & 0.491 & 0.066 & 13 \\
       & 54358.151 & 320.24 & 103.9 & 0.460 & 0.030 & 7 \\
       & 54358.157 & 319.77 & 105.3 & 0.415 & 0.034 & 8 \\
       & 54358.162 & 319.27 & 106.7 & 0.429 & 0.049 & 11 \\
145675 & 53958.259 & 329.81 & 168.4 & 0.902 & 0.054 & 6 \\
       & 53958.275 & 329.38 & 164.9 & 0.878 & 0.045 & 5 \\
       & 53958.292 & 328.62 & 161.0 & 0.859 & 0.051 & 6 \\
       & 53959.168 & 329.99 & 189.3 & 1.096 & 0.123 & 11 \\
       & 53959.184 & 330.18 & 185.5 & 1.000 & 0.089 & 9 \\
       & 53959.200 & 330.26 & 181.8 & 0.964 & 0.069 & 7 \\
       & 53959.215 & 330.26 & 178.1 & 0.990 & 0.070 & 7 \\
       & 53959.231 & 330.18 & 174.5 & 0.940 & 0.078 & 8 \\
       & 53959.246 & 330.01 & 170.9 & 0.954 & 0.067 & 7 \\
       & 53959.261 & 329.70 & 167.4 & 0.808 & 0.064 & 8 \\
154345 & 54719.168 & 328.79 & 90.5 & 0.885 & 0.094 & 11 \\
       & 54719.179 & 328.73 & 93.3 & 0.843 & 0.109 & 13 \\
       & 54719.185 & 328.66 & 94.7 & 0.811 & 0.089 & 11 \\
       & 54719.192 & 328.57 & 96.2 & 0.803 & 0.096 & 12 \\
       & 54719.198 & 328.45 & 97.6 & 0.847 & 0.096 & 11 \\
       & 54719.204 & 328.29 & 99.2 & 0.903 & 0.095 & 11 \\
       & 54719.213 & 328.00 & 101.4 & 0.817 & 0.122 & 15 \\
164922 & 54689.201 & 326.49 & 248.8 & 0.663 & 0.089 & 13 \\
       & 54689.212 & 325.27 & 251.2 & 0.820 & 0.058 & 7 \\
       & 54689.223 & 324.07 & 253.6 & 0.957 & 0.079 & 8 \\
       & 54689.235 & 322.84 & 256.4 & 1.067 & 0.153 & 14 \\
       & 54689.248 & 321.74 & 259.4 & 1.011 & 0.170 & 17 \\
167042 & 54358.232 & 321.20 & 97.5 & 0.584 & 0.037 & 6 \\
       & 54358.238 & 320.96 & 99.0 & 0.551 & 0.036 & 7 \\
       & 54358.243 & 320.68 & 100.3 & 0.507 & 0.036 & 7 \\
       & 54358.249 & 320.34 & 101.7 & 0.524 & 0.030 & 6 \\
       & 54358.255 & 319.96 & 103.1 & 0.571 & 0.036 & 6 \\
       & 54358.261 & 319.53 & 104.5 & 0.612 & 0.037 & 6 \\
       & 54358.267 & 319.05 & 105.9 & 0.591 & 0.041 & 7 \\
       & 54358.273 & 318.48 & 107.4 & 0.627 & 0.050 & 8 \\
170693 & 54346.303 & 187.40 & 183.8 & 0.373 & 0.042 & 11 \\
       & 54346.311 & 183.87 & 186.6 & 0.343 & 0.049 & 14 \\
       & 54346.321 & 179.32 & 190.2 & 0.358 & 0.037 & 10 \\
       & 54346.332 & 174.70 & 193.9 & 0.457 & 0.042 & 9 \\
185269 & 54665.204 & 321.00 & 228.6 & 0.860 & 0.146 & 17 \\
       & 54665.216 & 323.97 & 230.0 & 0.946 & 0.129 & 14 \\
       & 54665.226 & 326.17 & 231.3 & 0.757 & 0.148 & 20 \\
       & 54665.236 & 327.81 & 232.6 & 0.926 & 0.110 & 12 \\
       & 54665.245 & 328.96 & 233.9 & 0.928 & 0.178 & 19 \\
       & 54665.404 & 323.06 & 266.1 & 0.771 & 0.064 & 8 \\
       & 54665.410 & 322.92 & 267.7 & 0.741 & 0.050 & 7 \\
       & 54665.417 & 322.85 & 269.2 & 0.816 & 0.048 & 6 \\
       & 54665.423 & 322.85 & 90.8 & 0.921 & 0.057 & 6 \\
       & 54665.430 & 322.93 & 92.4 & 0.877 & 0.075 & 9 \\
       & 54665.438 & 323.11 & 94.3 & 0.912 & 0.084 & 9 \\
       & 54665.445 & 323.35 & 96.0 & 0.910 & 0.091 & 10 \\
       & 54665.452 & 323.68 & 97.7 & 0.855 & 0.080 & 9 \\
       & 54665.459 & 324.06 & 99.4 & 0.927 & 0.083 & 9 \\
       & 54665.466 & 324.52 & 101.1 & 0.841 & 0.129 & 15 \\
       & 54667.381 & 323.73 & 262.0 & 1.004 & 0.096 & 10 \\
       & 54667.387 & 323.44 & 263.5 & 0.830 & 0.103 & 12 \\
       & 54667.393 & 323.21 & 264.9 & 0.892 & 0.096 & 11 \\
       & 54667.400 & 323.02 & 266.5 & 1.014 & 0.085 & 8 \\
       & 54667.406 & 322.90 & 267.9 & 0.899 & 0.113 & 13 \\
188310 & 54717.211 & 293.54 & 249.5 & 0.103 & 0.014 & 14 \\
       & 54717.223 & 289.87 & 252.2 & 0.106 & 0.017 & 16 \\
       & 54717.229 & 288.10 & 253.7 & 0.107 & 0.012 & 11 \\
       & 54717.236 & 286.11 & 255.5 & 0.106 & 0.014 & 13 \\
       & 54717.242 & 284.72 & 257.0 & 0.094 & 0.015 & 16 \\
       & 54717.248 & 283.37 & 258.5 & 0.110 & 0.019 & 17 \\
       & 54717.253 & 282.29 & 260.0 & 0.111 & 0.018 & 16 \\
       & 54717.259 & 281.25 & 261.6 & 0.127 & 0.018 & 14 \\
199665 & 54717.336 & 285.96 & 90.6 & 0.614 & 0.064 & 10 \\
       & 54717.341 & 286.09 & 92.1 & 0.567 & 0.062 & 11 \\
       & 54717.347 & 286.36 & 93.6 & 0.562 & 0.077 & 14 \\
       & 54717.352 & 286.78 & 95.0 & 0.574 & 0.053 & 9 \\
       & 54717.358 & 287.37 & 96.6 & 0.566 & 0.060 & 11 \\
       & 54717.364 & 288.15 & 98.2 & 0.512 & 0.055 & 11 \\
       & 54717.370 & 289.11 & 99.1 & 0.479 & 0.069 & 14 \\
       & 54717.377 & 290.31 & 101.5 & 0.482 & 0.049 & 10 \\
       & 54717.383 & 291.58 & 103.1 & 0.414 & 0.035 & 8 \\
       & 54717.390 & 293.30 & 104.9 & 0.500 & 0.065 & 13 \\
210702 & 54717.426 & 302.96 & 100.6 & 0.635 & 0.076 & 12 \\
       & 54717.436 & 304.66 & 103.1 & 0.652 & 0.072 & 11 \\
       & 54717.442 & 305.68 & 104.5 & 0.591 & 0.085 & 14 \\
       & 54717.448 & 306.87 & 105.9 & 0.640 & 0.091 & 14 \\
217107 & 54717.283 & 292.41 & 236.3 & 0.771 & 0.096 & 12 \\
       & 54717.289 & 289.09 & 237.2 & 0.793 & 0.127 & 16 \\
       & 54717.296 & 285.35 & 238.3 & 0.757 & 0.095 & 13 \\
       & 54717.303 & 281.40 & 239.5 & 0.799 & 0.118 & 15 \\
       & 54717.309 & 278.11 & 240.6 & 0.776 & 0.114 & 15 \\
221345 & 54720.234 & 313.74 & 229.1 & 0.278 & 0.031 & 11 \\
       & 54720.239 & 315.41 & 229.9 & 0.253 & 0.034 & 13 \\
       & 54720.245 & 317.13 & 230.8 & 0.266 & 0.028 & 11 \\
       & 54720.250 & 318.64 & 231.7 & 0.232 & 0.024 & 10 \\
       & 54720.256 & 320.12 & 232.7 & 0.251 & 0.028 & 11 \\
222404 & 54664.457 & 253.07 & 230.4 & 0.105 & 0.011 & 10 \\
       & 54664.466 & 254.63 & 233.0 & 0.099 & 0.011 & 11 \\
       & 54664.475 & 256.07 & 235.6 & 0.091 & 0.010 & 11 \\
       & 54720.278 & 247.87 & 222.5 & 0.104 & 0.012 & 12 \\
       & 54720.285 & 249.26 & 224.5 & 0.093 & 0.010 & 11 \\
       & 54720.295 & 251.32 & 227.6 & 0.093 & 0.008 & 9 \\
       & 54720.301 & 252.45 & 229.3 & 0.086 & 0.008 & 9 \\
       & 54720.307 & 253.58 & 231.2 & 0.092 & 0.009 & 10 \\
       & 54720.313 & 254.70 & 233.2 & 0.091 & 0.008 & 9 \\
       & 54720.320 & 255.83 & 235.2 & 0.087 & 0.009 & 10 \\
\enddata
\tablecomments{The projected baseline position angle ($\Theta$) is calculated to be east of north.}
\end{deluxetable}

\clearpage

\begin{deluxetable}{ccccccccclc}
\tablewidth{1.1\textwidth}
\tabletypesize{\scriptsize}
\tablecaption{Exoplanet Host Star and Planet Observed Parameters \label{host_obs_props}}

\tablehead{ \multicolumn{1}{c}{ } & \multicolumn{5}{c}{Observed Stellar Parameters} & \multicolumn{1}{c}{ } &\multicolumn{3}{c}{Planetary System Parameters} \\
\cline{2-6} \cline{8-10}  \\ 

\colhead{ }  & \colhead{Spectral}  & \colhead{$\theta_{\rm LD}$} & \colhead{$\sigma_{\rm LD}$} & \colhead{$K$}   & \colhead{$\pi$} &  \colhead{ } & \colhead{$M_{\rm star}$} &  \colhead{$P$}     & \colhead{ }   & \colhead{ }  \\
\colhead{HD} & \colhead{Type}      & \colhead{(mas)}             & \colhead{$\%$}              & \colhead{(mag)} & \colhead{(mas)} &  \colhead{ } & \colhead{(M$_\odot$)}    & \colhead{(d)} & \colhead{Reference} & \colhead{(m-M)}\\ }
\startdata
10697 & G5 IV     & 0.49$\; \pm \;$0.05$^\mathrm{a}$ & 10 & 4.60$\; \pm \;$0.02 & 30.70$\; \pm \;$0.43 & & 1.2 & 1076.4 & \citet{2006ApJ...646..505B} & 2.6 \\
13189 & K2        & 0.84$\; \pm \;$0.03$^\mathrm{a}$ & 4  & 4.00$\; \pm \;$0.03 & 1.78$\; \pm \;$0.73   & & 3.5$^\mathrm{e}$ & 472 &  $P$ from \citet{2005AandA...437..743H} & 8.7 \\
32518 & K1 III    & 0.85$\; \pm \;$0.02$^\mathrm{b}$ & 2  & 3.91$\; \pm \;$0.04 & 8.29$\; \pm \;$0.58  & & 1.1 & 157.5  & \citet{2009AandA...505.1311D} & 5.4 \\
45410 & K0 III-IV & 0.97$\; \pm \;$0.04$^\mathrm{c}$ & 4  & 3.70$\; \pm \;$0.30 & 17.92$\; \pm \;$0.47 & & 1.7 & 889    & \citet{2008PASJ...60.1317S} & 3.7 \\
50554 & F8 V      & 0.34$\; \pm \;$0.10$^\mathrm{a}$ & 29 & 5.47$\; \pm \;$0.02 & 33.43$\; \pm \;$0.59 & & 1.1 & 1254   & \citet{2002PASP..114..529F} & 2.4 \\
73108 & K1 III    & 2.23$\; \pm \;$0.02$^\mathrm{b}$ & 1  & 1.92$\; \pm \;$0.07 & 12.74$\; \pm \;$0.26 & & 1.2 & 269.3  & \citet{2007AandA...472..649D} & 4.5 \\
136726 & K4 III   & 2.34$\; \pm \;$0.02$^\mathrm{b}$ & 1  & 1.92$\; \pm \;$0.05 & 8.19$\; \pm \;$0.19  & & 1.8 & 516.2  & \citet{2009AandA...505.1311D} & 5.4 \\
139357 & K4 III   & 1.07$\; \pm \;$0.01$^\mathrm{b}$ & 1  & 3.41$\; \pm \;$0.32 & 8.47$\; \pm \;$0.30  & & 1.3 & 1125.7 & \citet{2009AandA...499..935D} & 5.4 \\
145675 & K0 V     & 0.37$\; \pm \;$0.04$^\mathrm{a}$ & 11 & 4.71$\; \pm \;$0.02 & 56.91$\; \pm \;$0.34 & & 1.0 & 1724.0 & \citet{2003ApJ...582..455B} & 1.2 \\
154345 & G8 V     & 0.50$\; \pm \;$0.03$^\mathrm{c}$ & 6  & 5.00$\; \pm \;$0.02 & 53.80$\; \pm \;$0.32 & & 0.9 & 3360   & \citet{2008ApJ...683L..63W} & 1.3 \\
164922 & K0 V     & 0.50$\; \pm \;$0.07$^\mathrm{d}$ & 14 & 5.11$\; \pm \;$0.02 & 45.21$\; \pm \;$0.54 & & 0.9 & 1155   & \citet{2006ApJ...646..505B} & 1.7 \\
167042 & K1 III   & 0.92$\; \pm \;$0.02$^\mathrm{b}$ & 2  & 3.55$\; \pm \;$0.24 & 19.91$\; \pm \;$0.26 & & 1.5 & 418    & \citet{2008PASJ...60.1317S} & 3.5 \\
170693 & K1.5 III & 2.04$\; \pm \;$0.04$^\mathrm{b}$ & 2  & 1.95$\; \pm \;$0.05 & 10.36$\; \pm \;$0.20 & & 1.0 & 479.1  & \citet{2009AandA...499..935D} & 4.9 \\
185269 & G0 IV    & 0.48$\; \pm \;$0.03$^\mathrm{c}$ & 6  & 5.26$\; \pm \;$0.02 & 19.89$\; \pm \;$0.56 & & 1.3 & 6.8    & \citet{2006ApJ...652.1724J} & 3.5 \\
188310 & G9 III   & 1.73$\; \pm \;$0.01$^\mathrm{c}$ & 1  & 2.17$\; \pm \;$0.22 & 17.77$\; \pm \;$0.29 & & 2.2 & 137    & \citet{2008PASJ...60..539S} & 3.8 \\
199665 & G6 III   & 1.11$\; \pm \;$0.03$^\mathrm{c}$ & 3  & 3.37$\; \pm \;$0.20 & 13.28$\; \pm \;$0.31 & & 2.2 & 993    & \citet{2008PASJ...60..539S} & 4.4 \\
210702 & K1 III   & 0.88$\; \pm \;$0.02$^\mathrm{c}$ & 2  & 3.98$\; \pm \;$0.29 & 18.20$\; \pm \;$0.39 & & 1.9 & 341.1  & \citet{2007ApJ...665..785J} & 3.7 \\
217107 & G8 IV    & 0.70$\; \pm \;$0.01$^\mathrm{c}$ & 1  & 4.54$\; \pm \;$0.02 & 50.36$\; \pm \;$0.38 & & 1.0 & 7.1    & \citet{1999PASP..111...50F} & 1.5 \\
221345 & G8 III   & 1.34$\; \pm \;$0.01$^\mathrm{c}$ & 1  & 2.33$\; \pm \;$0.24 & 12.63$\; \pm \;$0.27 & & 2.2 & 186    & \citet{2008PASJ...60.1317S} & 4.5 \\
222404 & K1 IV    & 3.30$\; \pm \;$0.03$^\mathrm{c}$ & 1  & 1.04$\; \pm \;$0.21 & 70.91$\; \pm \;$0.40 & & 1.6 & 906    & \citet{2003ApJ...599.1383H} & 0.7 \\

\enddata
\tablecomments{Spectral types are from \emph{SIMBAD}; parallaxes $\pi$ are from \citet{2007hnrr.book.....V}; $K$ magnitudes are from \citet{2003tmc..book.....C}, except for HD~73108, HD~136726, and HD~170693, which are from \citet{1969tmss.book.....N}; (m-M) was calculated using Equation 3; $^\mathrm{a}$\citet{2008ApJ...680..728B}; $^\mathrm{b}$\citet{2010ApJ...710.1365B}; $^\mathrm{c}$\citet{2009ApJ...701..154B}; $^\mathrm{d}$previously unpublished; $^\mathrm{e}$Mass from \citet{2005ApJ...632L.131S}, though they cannot constrain the mass to better than 2-6$M_\odot$}
\end{deluxetable}

\clearpage

\begin{deluxetable}{ccccccccccccccccccccc}
\rotate
\tablewidth{1.4\textwidth}
\tabletypesize{\scriptsize}
\tablecaption{Calculated Parameters for Secondary Stars of Various Spectral Types \label{poss_sec_params}}

\tablehead{\multicolumn{1}{c}{ } & \multicolumn{1}{c}{ } & \multicolumn{3}{c}{G5 V} & \multicolumn{1}{c}{ } & \multicolumn{3}{c}{K0 V} & \multicolumn{1}{c}{ } & \multicolumn{3}{c}{K5 V} & \multicolumn{1}{c}{ } & \multicolumn{3}{c}{M0 V} & \multicolumn{1}{c}{ } & \multicolumn{3}{c}{M5 V} \\
\cline{3-5} \cline{7-9} \cline{11-13} \cline{15-17} \cline{19-21}  \\ 
\colhead{ }  & \colhead{ } & \colhead{$\Delta K$} & \colhead{$\alpha$} & \colhead{$\theta$} & \colhead{ } & \colhead{$\Delta K$} & \colhead{$\alpha$} & \colhead{$\theta$} & \colhead{ } & \colhead{$\Delta K$} & \colhead{$\alpha$} & \colhead{$\theta$} & \colhead{ } & \colhead{$\Delta K$} & \colhead{$\alpha$} & \colhead{$\theta$} & \colhead{ } & \colhead{$\Delta K$} & \colhead{$\alpha$} & \colhead{$\theta$} \\
\colhead{HD} & \colhead{ } & \colhead{(mag)}      & \colhead{(mas)}    & \colhead{(mas)}    & \colhead{ } & \colhead{(mag)}      & \colhead{(mas)}    & \colhead{(mas)}    & \colhead{ } & \colhead{(mag)}      & \colhead{(mas)}    & \colhead{(mas)}    & \colhead{ } & \colhead{(mag)}      & \colhead{(mas)}    & \colhead{(mas)}    & \colhead{ } & \colhead{(mag)}      & \colhead{(mas)}    & \colhead{(mas)} \\ }
\startdata
10697 &  & 1.5 & 80.6 & 0.26 &  & 1.9 & 78.8 & 0.24 &  & 2.5 & 77.2 & 0.21 &  & 3.1 & 74.9 & 0.17 &  & 4.1 & 70.1 & 0.08 \\
13189 &  & 8.3 & 3.5 & 0.02 &  & 8.7 & 3.4 & 0.01 &  & 9.3 & 3.4 & 0.01 &  & 9.9 & 3.4 & 0.01 &  & 10.9 & 3.3 & 0.00 \\
32518 &  & 5.0 & 6.0 & 0.07 &  & 5.4 & 5.9 & 0.07 &  & 6.0 & 5.8 & 0.06 &  & 6.6 & 5.6 & 0.05 &  & 7.6 & 5.2 & 0.02 \\
45410 &  & 3.5 & 44.7 & 0.15 &  & 3.9 & 43.9 & 0.14 &  & 4.5 & 43.2 & 0.12 &  & 5.2 & 42.2 & 0.10 &  & 6.2 & 40.2 & 0.05 \\
50554 &  & 0.4 & 95.6 & 0.29 &  & 0.8 & 93.5 & 0.26 &  & 1.4 & 91.4 & 0.22 &  & 2.1 & 88.5 & 0.19 &  & 3.0 & 82.5 & 0.08 \\
73108 &  & 6.1 & 13.4 & 0.11 &  & 6.5 & 13.1 & 0.10 &  & 7.1 & 12.9 & 0.09 &  & 7.7 & 12.5 & 0.07 &  & 8.7 & 11.7 & 0.03 \\
136726 &  & 7.0 & 14.4 & 0.07 &  & 7.4 & 14.2 & 0.06 &  & 8.0 & 13.9 & 0.05 &  & 8.7 & 13.6 & 0.05 &  & 9.6 & 13.0 & 0.02 \\
139357 &  & 5.5 & 23.4 & 0.07 &  & 5.9 & 23.0 & 0.07 &  & 6.5 & 22.5 & 0.06 &  & 7.1 & 21.9 & 0.05 &  & 8.1 & 20.6 & 0.02 \\
145675 &  & 0.0 & 199.0 & 0.49 &  & 0.4 & 194.4 & 0.45 &  & 1.0 & 190.0 & 0.38 &  & 1.7 & 183.7 & 0.32 &  & 2.6 & 170.6 & 0.14 \\
154345 &  & -0.2 & 287.3 & 0.46 &  & 0.2 & 280.3 & 0.43 &  & 0.8 & 273.4 & 0.36 &  & 1.5 & 263.6 & 0.30 &  & 2.5 & 243.1 & 0.14 \\
164922 &  & 0.1 & 119.8 & 0.39 &  & 0.5 & 116.9 & 0.36 &  & 1.1 & 114.2 & 0.30 &  & 1.8 & 110.2 & 0.25 &  & 2.7 & 102.0 & 0.11 \\
167042 &  & 3.5 & 29.2 & 0.17 &  & 3.9 & 28.7 & 0.16 &  & 4.5 & 28.2 & 0.13 &  & 5.1 & 27.5 & 0.11 &  & 6.1 & 26.0 & 0.05 \\
170693 &  & 6.5 & 15.4 & 0.09 &  & 6.9 & 15.0 & 0.08 &  & 7.5 & 14.7 & 0.07 &  & 8.1 & 14.2 & 0.06 &  & 9.1 & 13.2 & 0.03 \\
185269 &  & 1.7 & 1.8 & 0.17 &  & 2.1 & 1.8 & 0.16 &  & 2.7 & 1.7 & 0.13 &  & 3.4 & 1.7 & 0.11 &  & 4.4 & 1.6 & 0.05 \\
188310 &  & 5.1 & 13.5 & 0.15 &  & 5.5 & 13.3 & 0.14 &  & 6.1 & 13.1 & 0.12 &  & 6.7 & 12.9 & 0.10 &  & 7.7 & 12.4 & 0.04 \\
199665 &  & 4.5 & 37.8 & 0.11 &  & 4.9 & 37.3 & 0.11 &  & 5.5 & 36.8 & 0.09 &  & 6.2 & 36.1 & 0.07 &  & 7.1 & 34.7 & 0.03 \\
210702 &  & 3.2 & 24.4 & 0.16 &  & 3.6 & 24.0 & 0.14 &  & 4.2 & 23.7 & 0.12 &  & 4.9 & 23.2 & 0.10 &  & 5.8 & 22.1 & 0.05 \\
217107 &  & 0.5 & 4.5 & 0.43 &  & 0.9 & 4.4 & 0.40 &  & 1.5 & 4.3 & 0.34 &  & 2.1 & 4.1 & 0.28 &  & 3.1 & 3.8 & 0.13 \\
221345 &  & 5.7 & 11.8 & 0.11 &  & 6.1 & 11.6 & 0.10 &  & 6.7 & 11.4 & 0.08 &  & 7.3 & 11.2 & 0.07 &  & 8.3 & 10.8 & 0.03 \\
222404 &  & 3.2 & 176.6 & 0.61 &  & 3.6 & 173.5 & 0.56 &  & 4.2 & 170.5 & 0.48 &  & 4.9 & 166.4 & 0.40 &  & 5.8 & 158.1 & 0.18 \\

\enddata
\tablecomments{Values for M$_K$ (used to calculate $\Delta K$), secondary stellar masses (used to calculate $\alpha$), and secondary stellar radii (used to calculate $\theta$) were obtained from \citet{2000asqu.book.....C}: G5~V~= 3.5, 0.92~$M_\odot$, 0.92~$R_\odot$; K0~V~=~3.9, 0.79~$M_\odot$, 0.85~$R_\odot$; K5~V~=~4.5, 0.67~$M_\odot$, 0.72~$R_\odot$; M0~V~=~5.2, 0.51~$M_\odot$, 0.60~$R_\odot$; M5~V~=~6.1, 0.21~$M_\odot$, 0.27~$R_\odot$.}
\end{deluxetable}

\clearpage

\begin{deluxetable}{clccccccccccccccccccc}
\rotate
\tablewidth{1.4\textwidth}
\tabletypesize{\scriptsize}
\tablecaption{Observed Diameter Fit Residuals and Calculated Binary Visibility Residuals \label{calc_resids}}

\tablehead{\multicolumn{4}{c}{ } & \multicolumn{5}{c}{$\Delta V_{\rm max}$, PA=0$^\circ$} & \multicolumn{1}{c}{ } & \multicolumn{5}{c}{$\Delta V_{\rm max}$, PA=30$^\circ$} & \multicolumn{1}{c}{ } & \multicolumn{5}{c}{$\Delta V_{\rm max}$, PA=60$^\circ$} \\ 
 \cline{5-9} \cline{11-15} \cline{17-21} \\ 
 \colhead{HD} & \colhead{Obs Date} & \colhead{$\sigma_{\rm res}$} & \colhead{ } & \colhead{G5V} & \colhead{K0V} & \colhead{K5V} & \colhead{M0V} & \colhead{M5V} & \colhead{ } & \colhead{G5V} & \colhead{K0V} & \colhead{K5V} & \colhead{M0V} & \colhead{M5V} & \colhead{ } & \colhead{G5V} & \colhead{K0V} & \colhead{K5V} & \colhead{M0V} &  \colhead{M5V} }
\startdata
10697  & 2005 Oct 23 & 0.092 &  & 0.221 & 0.162 & 0.098 & 0.058 & 0.024 &  & 0.232 & 0.164 & 0.096 & 0.055 & 0.022 &  & 0.225 & 0.162 & 0.098 & 0.058 & 0.024 \\
       & 2007 Sep 14 & 0.053 &  & 0.221 & 0.162 & 0.098 & 0.058 & 0.024 &  & 0.232 & 0.164 & 0.096 & 0.055 & 0.022 &  & 0.225 & 0.162 & 0.098 & 0.058 & 0.024 \\
13189  & 2005 Dec 12 & 0.033 &  & 0.001 & 0.000 & 0.000 & 0.000 & 0.000 &  & 0.001 & 0.000 & 0.000 & 0.000 & 0.000 &  & 0.001 & 0.000 & 0.000 & 0.000 & 0.000 \\
       & 2006 Aug 14 & 0.019 &  & 0.001 & 0.000 & 0.000 & 0.000 & 0.000 &  & 0.001 & 0.000 & 0.000 & 0.000 & 0.000 &  & 0.001 & 0.000 & 0.000 & 0.000 & 0.000 \\
32518  & 2007 Nov 14 & 0.036 &  & 0.011 & 0.008 & 0.005 & 0.003 & 0.001 &  & 0.012 & 0.008 & 0.005 & 0.003 & 0.001 &  & 0.011 & 0.008 & 0.005 & 0.003 & 0.001 \\
45410  & 2008 Sep 11 & 0.052 &  & 0.047 & 0.033 & 0.019 & 0.010 & 0.004 &  & 0.046 & 0.032 & 0.019 & 0.010 & 0.004 &  & 0.047 & 0.033 & 0.019 & 0.010 & 0.004 \\
50554  & 2005 Dec 12 & 0.047 &  & 0.448 & 0.346 & 0.213 & 0.112 & 0.057 &  & 0.445 & 0.365 & 0.238 & 0.135 & 0.062 &  & 0.485 & 0.384 & 0.244 & 0.137 & 0.061 \\
73108  & 2008 May 09 & 0.018 &  & 0.200 & 0.201 & 0.202 & 0.202 & 0.202 &  & 0.201 & 0.201 & 0.201 & 0.202 & 0.202 &  & 0.199 & 0.201 & 0.202 & 0.202 & 0.202 \\
136726 & 2008 May 09 & 0.018 &  & 0.194 & 0.194 & 0.194 & 0.194 & 0.195 &  & 0.194 & 0.194 & 0.194 & 0.194 & 0.195 &  & 0.193 & 0.194 & 0.194 & 0.194 & 0.195 \\
139357 & 2007 Sep 14 & 0.019 &  & 0.008 & 0.005 & 0.003 & 0.002 & 0.001 &  & 0.008 & 0.005 & 0.003 & 0.002 & 0.001 &  & 0.008 & 0.005 & 0.003 & 0.002 & 0.001 \\
145675 & 2006 Aug 12 & 0.056 &  & 0.543 & 0.460 & 0.302 & 0.166 & 0.076 &  & 0.401 & 0.420 & 0.301 & 0.165 & 0.046 &  & 0.542 & 0.459 & 0.302 & 0.169 & 0.077 \\
154345 & 2008 Sep 10 & 0.039 &  & 0.256 & 0.336 & 0.282 & 0.186 & 0.094 &  & 0.493 & 0.529 & 0.391 & 0.233 & 0.099 &  & 0.272 & 0.361 & 0.328 & 0.211 & 0.094 \\
164922 & 2008 Aug 11 & 0.161 &  & 0.564 & 0.489 & 0.322 & 0.183 & 0.082 &  & 0.608 & 0.481 & 0.306 & 0.167 & 0.069 &  & 0.562 & 0.487 & 0.321 & 0.183 & 0.082 \\
167042 & 2007 Sep 15 & 0.040 &  & 0.045 & 0.032 & 0.019 & 0.011 & 0.004 &  & 0.046 & 0.032 & 0.019 & 0.011 & 0.004 &  & 0.046 & 0.032 & 0.018 & 0.011 & 0.004 \\
170693 & 2007 Sep 03 & 0.037 &  & 0.219 & 0.219 & 0.220 & 0.220 & 0.220 &  & 0.219 & 0.219 & 0.220 & 0.220 & 0.220 &  & 0.218 & 0.219 & 0.220 & 0.220 & 0.220 \\
185269 & 2008 Jul 18 & 0.078 &  & 0.193 & 0.139 & 0.086 & 0.047 & 0.019 &  & 0.203 & 0.146 & 0.087 & 0.047 & 0.018 &  & 0.137 & 0.098 & 0.058 & 0.032 & 0.013 \\
       & 2008 Jul 20 & 0.079 &  & 0.193 & 0.139 & 0.086 & 0.047 & 0.019 &  & 0.203 & 0.146 & 0.087 & 0.047 & 0.018 &  & 0.137 & 0.098 & 0.058 & 0.032 & 0.013 \\
188310 & 2008 Sep 08 & 0.012 &  & 0.151 & 0.152 & 0.153 & 0.154 & 0.155 &  & 0.151 & 0.152 & 0.153 & 0.154 & 0.155 &  & 0.152 & 0.152 & 0.153 & 0.154 & 0.155 \\
199665 & 2008 Sep 08 & 0.054 &  & 0.020 & 0.014 & 0.008 & 0.004 & 0.002 &  & 0.020 & 0.014 & 0.008 & 0.004 & 0.002 &  & 0.020 & 0.014 & 0.008 & 0.004 & 0.002 \\
210702 & 2008 Sep 08 & 0.025 &  & 0.060 & 0.042 & 0.024 & 0.013 & 0.006 &  & 0.059 & 0.041 & 0.024 & 0.013 & 0.006 &  & 0.059 & 0.041 & 0.024 & 0.013 & 0.006 \\
217107 & 2008 Sep 08 & 0.017 &  & 0.496 & 0.376 & 0.239 & 0.148 & 0.061 &  & 0.499 & 0.374 & 0.234 & 0.140 & 0.059 &  & 0.454 & 0.339 & 0.213 & 0.131 & 0.058 \\
221345 & 2008 Sep 11 & 0.013 &  & 0.007 & 0.005 & 0.003 & 0.002 & 0.001 &  & 0.007 & 0.005 & 0.003 & 0.002 & 0.001 &  & 0.007 & 0.005 & 0.003 & 0.002 & 0.001 \\
222404 & 2008 Sep 11 & 0.006 &  & 0.182 & 0.187 & 0.192 & 0.196 & 0.198 &  & 0.182 & 0.187 & 0.192 & 0.196 & 0.198 &  & 0.182 & 0.187 & 0.192 & 0.196 & 0.198 \\
\enddata
\tablecomments{PA is the position angle.}
\end{deluxetable}

\clearpage

\begin{deluxetable}{clcccccccccccccccccc}
\rotate
\tablewidth{1.35\textwidth}
\tabletypesize{\scriptsize}
\tablecaption{Eliminating Certain Secondary Stars \label{xp_resids}}

\tablehead{\multicolumn{3}{c}{ } & \multicolumn{5}{c}{PA=0$^\circ$} & \multicolumn{1}{c}{ } & \multicolumn{5}{c}{PA=30$^\circ$} & \multicolumn{1}{c}{ } & \multicolumn{5}{c}{PA=60$^\circ$} \\ 
 \cline{4-8} \cline{10-14} \cline{16-20} \\ 
 \colhead{HD} & \colhead{Obs Date} & \colhead{ } & \colhead{G5V} & \colhead{K0V} & \colhead{K5V} & \colhead{M0V} & \colhead{M5V} & \colhead{ } & \colhead{G5V} & \colhead{K0V} & \colhead{K5V} & \colhead{M0V} & \colhead{M5V} & \colhead{ } & \colhead{G5V} & \colhead{K0V} & \colhead{K5V} & \colhead{M0V} &  \colhead{M5V} }
\startdata
10697  & 2005 Oct 23 &  & X & P & P & P & P &  & X & P & P & P & P &  & X & P & P & P & P \\
       & 2007 Sep 14 &  & X & X & P & P & P &  & X & X & P & P & P &  & X & X & P & P & P \\
13189  & 2005 Dec 12 &  & P & P & P & P & P &  & P & P & P & P & P &  & P & P & P & P & P \\
       & 2006 Aug 14 &  & P & P & P & P & P &  & P & P & P & P & P &  & P & P & P & P & P \\
32518  & 2007 Nov 14 &  & P & P & P & P & P &  & P & P & P & P & P &  & P & P & P & P & P \\
45410  & 2008 Sep 11 &  & P & P & P & P & P &  & P & P & P & P & P &  & P & P & P & P & P \\
50554  & 2005 Dec 12 &  & X & X & X & X & P &  & X & X & X & X & P &  & X & X & X & X & P \\
73108  & 2008 May 09 &  & X & X & X & X & X &  & X & X & X & X & X &  & X & X & X & X & X \\
136726 & 2008 May 09 &  & X & X & X & X & X &  & X & X & X & X & X &  & X & X & X & X & X \\
139357 & 2007 Sep 14 &  & P & P & P & P & P &  & P & P & P & P & P &  & P & P & P & P & P \\
145675 & 2006 Aug 12 &  & X & X & X & X & P &  & X & X & X & X & P &  & X & X & X & X & P \\
154345 & 2008 Sep 10 &  & X & X & X & X & X &  & X & X & X & X & X &  & X & X & X & X & X \\
164922 & 2008 Aug 11 &  & X & X & P & P & P &  & X & X & P & P & P &  & X & X & P & P & P \\
167042 & 2007 Sep 15 &  & P & P & P & P & P &  & P & P & P & P & P &  & P & P & P & P & P \\
170693 & 2007 Sep 03 &  & X & X & X & X & X &  & X & X & X & X & X &  & X & X & X & X & X \\
185269 & 2008 Jul 18 &  & X & P & P & P & P &  & X & P & P & P & P &  & P & P & P & P & P \\
       & 2008 Jul 20 &  & X & P & P & P & P &  & X & P & P & P & P &  & P & P & P & P & P \\
188310 & 2008 Sep 08 &  & X & X & X & X & X &  & X & X & X & X & X &  & X & X & X & X & X \\
199665 & 2008 Sep 08 &  & P & P & P & P & P &  & P & P & P & P & P &  & P & P & P & P & P \\
210702 & 2008 Sep 08 &  & X & P & P & P & P &  & X & P & P & P & P &  & X & P & P & P & P \\
217107 & 2008 Sep 08 &  & X & X & X & X & X &  & X & X & X & X & X &  & X & X & X & X & X \\
221345 & 2008 Sep 11 &  & P & P & P & P & P &  & P & P & P & P & P &  & P & P & P & P & P \\
222404 & 2008 Sep 11 &  & X & X & X & X & X &  & X & X & X & X & X &  & X & X & X & X & X \\

\enddata
\tablecomments{``X'' indicates a secondary star of that spectral type can be ruled out from the observations, while ``P'' means a secondary star with that spectral type is still a possibility in the system.}
\end{deluxetable}

\clearpage

\begin{deluxetable}{cccc}
\tablewidth{0pc}
\tabletypesize{\scriptsize}
\tablecaption{Angular Diameter Comparison \label{ld_sed_vk}}

\tablehead{ 

\colhead{ }  & \colhead{$\theta_{\rm LD}$} & \colhead{$\theta_{\rm SED}$} & \colhead{$\theta_{(V-K)}$}  \\
\colhead{HD} & \colhead{(mas)}             & \colhead{(mas)}              & \colhead{(mas)}}
\startdata
10697  & 0.49$\; \pm \;$0.05 & 0.50$\; \pm \;$0.03 & 0.53$\; \pm \;$0.02  \\
13189  & 0.84$\; \pm \;$0.03 & 0.80$\; \pm \;$0.04 & 0.97$\; \pm \;$0.04  \\
32518  & 0.85$\; \pm \;$0.02 & 0.89$\; \pm \;$0.10 & 0.84$\; \pm \;$0.04 \\
45410  & 0.97$\; \pm \;$0.04 & 0.87$\; \pm \;$0.07 & 0.87$\; \pm \;$0.37  \\
50554  & 0.34$\; \pm \;$0.10 & 0.33$\; \pm \;$0.01 & 0.34$\; \pm \;$0.01 \\
73108  & 2.23$\; \pm \;$0.02 & 2.12$\; \pm \;$0.15 & 2.44$\; \pm \;$0.75 \\
136726 & 2.34$\; \pm \;$0.02 & 2.46$\; \pm \;$0.34 & 2.30$\; \pm \;$0.88 \\
139357 & 1.07$\; \pm \;$0.01 & 1.40$\; \pm \;$0.28 & 1.07$\; \pm \;$0.49 \\
145675 & 0.37$\; \pm \;$0.04 & 0.51$\; \pm \;$0.01 & 0.53$\; \pm \;$0.01 \\
154345 & 0.50$\; \pm \;$0.03 & 0.43$\; \pm \;$0.02 & 0.45$\; \pm \;$0.01  \\
164922 & 0.50$\; \pm \;$0.07 & 0.39$\; \pm \;$0.02 & 0.44$\; \pm \;$0.02 \\
167042 & 0.92$\; \pm \;$0.02 & 0.88$\; \pm \;$0.07 & 0.98$\; \pm \;$0.33 \\
170693 & 2.04$\; \pm \;$0.04 & 1.90$\; \pm \;$0.16 & 2.03$\; \pm \;$0.62 \\
185269 & 0.48$\; \pm \;$0.03 & 0.34$\; \pm \;$0.02 & 0.38$\; \pm \;$0.01  \\
188310 & 1.73$\; \pm \;$0.01 & 1.68$\; \pm \;$0.06 & 1.88$\; \pm \;$0.59  \\
199665 & 1.11$\; \pm \;$0.03 & 0.99$\; \pm \;$0.03 & 1.01$\; \pm \;$0.29 \\
210702 & 0.88$\; \pm \;$0.02 & 0.83$\; \pm \;$0.07 & 0.74$\; \pm \;$0.31  \\
217107 & 0.70$\; \pm \;$0.01 & 0.52$\; \pm \;$0.02 & 0.54$\; \pm \;$0.02 \\
221345 & 1.34$\; \pm \;$0.01 & 1.43$\; \pm \;$0.14 & 1.86$\; \pm \;$0.63  \\
222404 & 3.30$\; \pm \;$0.03 & 3.06$\; \pm \;$0.16 & 2.98$\; \pm \;$0.87  \\
\enddata
\tablecomments{For the SED and ($V-K$) fits, the photometric values used are from the following sources: $UBV$ from \citet{Mermilliod} for all stars except HD~13189 and HD~50554 \citep[$BV$ only from ][]{1997yCat.1239....0E}; $RI$ from \citet{2003AJ....125..984M}; and $JHK$ from \citet{2003tmc..book.....C}. $T_{\rm eff}$ and log~$g$ values were from \citet{1999A&A...352..555A} for all stars except HD~13189 and HD~50554 \citep{2010arXiv1004.1069S}; HD~136726 \citep{1997A&AS..124..299C}; HD~154345 \citep{2007yCat.3251....0P}; and HD~32518 and HD~139357, which are from \citet{2000asqu.book.....C} and were based on their spectral types as listed in the \emph{SIMBAD Astronomical Database}.}
\end{deluxetable}

\clearpage


\clearpage

\clearpage

\clearpage

\end{document}